\documentclass[aps,prl,preprint,groupedaddress]{revtex4-1}

\usepackage{graphicx}
\usepackage{dcolumn}
\usepackage{amsmath}
\usepackage{color}

\def\be{\begin{equation}}
\def\ee{\end{equation}}
\def\beq{\begin{eqnarray}}
\def\eeq{\end{eqnarray}}
\def\n{\nonumber}

\begin{document}


\title{ Compact objects in pure Lovelock theory}


\author{Naresh Dadhich}  \email[]{nkd@iucaa.in}
\affiliation{ Centre for Theoretical Physics, Jamia Millia, Islamia, New Delhi, 110025, India\\ Inter-University Center for Astronomy and Astrophysics, Post Bag 4 Pune 411 007, India \\
 Astrophysics and Cosmology Research Unit,
 School of Mathematics, Statistics and Computer Science, University of KwaZulu-Natal,  Private Bag X54001, Durban 4000, South Africa}

\author{Sudan Hansraj} \email[]{hansrajs@ukzn.ac.za}
\affiliation{ Astrophysics and Cosmology Research Unit, School of Mathematics, Statistics and Computer Science,
University of KwaZulu-Natal, Private Bag X54001, Durban 4000, South Africa}

\author{Brian Chilambwe} \email[]{brian@aims.ac.za}
\affiliation{ Astrophysics and Cosmology Research Unit, School of Mathematics, Statistics and Computer Science,
University of KwaZulu-Natal, Private Bag X54001, Durban 4000, South Africa}


\date{\today}

\begin{abstract}
For static fluid interiors of compact objects in pure Lovelock gravity (involving only one
$N$th order term in the equation) we establish similarity in solutions for the critical
odd and even $d=2N+1, 2N+2$ dimensions. It turns out that in critical odd $d=2N+1$ dimensions, there
cannot exist any bound distribution with a finite radius, while in critical even
$d=2N+2$ dimensions, all solutions have similar behavior. For
exhibition of similarity we would compare star solutions for $N =1, 2$ in
$d=4$ Einstein and $d=6$ in Gauss-Bonnet theory respectively. We also
obtain the pure Lovelock analogue of the Finch-Skea model.

\end{abstract}




\maketitle


\section{Introduction}

The most natural and elegant generalization of Einstein gravity is Lovelock gravity which involves
in action a homogeneous polynomial of degree $N$ in Riemann curvature. Despite this, it yet yields a
second order equation of motion. This is the unique Lovelock \cite{lov} property which no other generalization
can boast of. It includes corresponding to $N=0, 1, 2, ...$ respectively the cosmological constant,
Einstein and Gauss--Bonnet (GB) gravity, and so on. It is also remarkable that higher order terms make
non-zero contribution in the equation only for dimensions $d \geq2N+1$. It is therefore an essential
higher dimensional generalization of Einstein gravity. \\

Gravitational solutions for vacuum as well as for fluid spheres have been studied for the Lovelock gravity by
several authors. Beginning with the Einstein-GB (EGB) black hole solution \cite{bd}, vacuum solutions have been obtained for Lovelock gravity in general \cite{whit, whee} as well as for dimensionally continued \cite{Ban1,Ban2,Myers,Wilt}, and also for pure Lovelock black holes \cite{probes3,cai}. For the latter, the vacuum solutions are non-trivial only for $d\geq2N+2$. Polytropic equations of state for isothermal fluid spheres were examined by Tooper \cite{Toop1, Toop2,Toop3, Toop4} in standard Einstein gravity theory.  The configuration of a differentially rotating isothermal fluid sphere was discussed by Hartle \cite{Hart1, Hart2, Hart3}. It was argued by Sharma \cite{Sharm} that the limiting case of the polytrope may be regarded as isothermal and is important in the study of clusters of stars.
Some fluid interior stellar models have also been obtained in EGB gravity \cite{dkm,maha-hans,hans-maha,chil-hans}, and in particular it is shown \cite{dkm} that the Schwarzschild interior solution is universal in describing a uniform density sphere in all dimensions $>3$ irrespective of the gravitational theory, Einstein or Lovelock.

It is well known that Einstein gravity is kinematic, implying no non-trivial vacuum solution, in $3$ dimensions,
and it becomes dynamic in $4$ dimensions. It turns out that this property can be universalized for all odd dimensions $d=2N+1$ dimensions and it then uniquely picks out pure Lovelock equation in higher dimensions \cite{nd, nd2}. This is done by defining $N$th order Lovelock Riemann curvature which is a $N$th degree homogeneous polynomial in Riemann \cite{bianchi, cam-d}. Then in all critical odd dimensions $d=2N+1$, the Lovelock  Riemann is entirely given in terms of the corresponding Ricci \cite{cam-d} thereby implying the non-existence of non-trivial pure Lovelock vacuum solution \cite{bound, cam-d}. It has been strongly argued by one of us  \cite{bound, nd2} that the pure Lovelock equation is the right equation in higher dimensions and it is appropriate only in two odd and even dimensions  $d=2N+1, 2N+2$. Gravity has qualitatively similar behaviour in these critical dimensions, i.e. what Einstein gravity is in $3, 4$ dimensions, so is pure GB in $5, 6$ dimensions, and so on. For a given dimension, $N=(d-1)/2$ is the degree of Riemann involved in action and equation of motion. Generally speaking gravitational  dynamics in the critical dimensions $d=2N+1, 2N+2$   have a universal character in pure Lovelock gravity. \\

Note that each term in the Lovelock Lagrangian comes with a dimensionful coupling which cannot all be determined because there is only one gravitational force that can fix only one coupling constant by measuring experimentally the strength of the force. What string or brane gravity theorists do is that all higher order couplings are expanded in inverse powers  of Newton's constant $G$. This is fine when one is considering high energy corrections to the Einstein theory. Here we are on the other hand considering a classical gravity in higher dimensions, and not a modification to the Einstein gravity. This is why it becomes  pertinent to ask  what is the correct classical equation for gravity in higher dimensions? Apart from the universalization of kinematic property to all critical odd dimensions, there is an even more compelling motivation that bound orbits should exist around a static object. Note that for Einstein gravity bound orbits exist only in four  dimensions. Existence of bound orbits again uniquely pick out pure Lovelock gravity [30]. Thus pure Lovelock gravity is the proper gravitational theory in higher dimensions.

It is therefore important to verify universality in various situations. Besides the uniform density sphere being always described \cite{dkm} by the Schwarzschild interior solution, very recently it has been shown that the pure Lovelock isothermal fluid sphere is also universal \cite{isothermal} in all dimensions $ d \geq 2N+2$. We would like to carry it forward by studying pure Lovelock solutions for a static fluid sphere. For a fluid sphere, the only equation to be solved is that of pressure isotropy. Note that in classical general relativity the TOV equations \cite{T,OV} have played a vital role in studying exact solutions  and in understanding the internal structure of stars \cite{Misner}.  It turns out that it has the same character in critical odd and even dimensions $d=2N+1, 2N+2$. Since pure Lovelock gravity is kinematic in critical odd dimensions $d=2N+1$  it means that there cannot exist  any non-trivial vacuum solution which can be an exterior to a fluid distribution. This implies that in the in the critical odd dimensions, there cannot occur any bound fluid distribution. There is only one equation to determine two metric functions. This cannot be done unless either one of the functions is prescribed or there is an equation of state relating density and pressure or a fall off behavior for density is prescribed. \\

There are very few fully physically viable solutions, the one due to Finch and Skea \cite{Finch-Skea} is particularly interesting for its physical viability and acceptability \cite{wal}. We shall first obtain pure Lovelock analogue of the Finch-Skea solution and in particular show that GB solution in $d=5, 6$ is similar to the Einstein solution in $d=3, 4$. Very recently a general Buchdahl \cite{buch} ansatz for a static fluid sphere has been studied
for Einstein and pure Lovelock gravity \cite{kdm, mdk} and it includes all the physically tenable models including Vaidya-Tikekar \cite{vt} and Finch-Skea \cite{Finch-Skea}. There, the authors observe that the pressure isotropy equation has the same form for a given Lovelock order $N$ in all dimensions, and hence a solution could be lifted from a lower  to a higher dimension. Here our concern is to establish similarity between the solutions in critical odd and even dimensions for two different values of the Lovelock order $N$.

The paper is organised as follows. In the next section we set up the pure Lovelock equation of motion which in the following section is  specialized to a spherically symmetric spacetime for the study of fluid sphere solutions. The specific case $d = 2N + 1$ is immediately analysed and it is shown that bounded distributions cannot exist with this prescription. The Schwarzschild and Finch-Skea metrics are presented as examples. Thereafter the assumption of constant density is shown to be a necessary and sufficient condition for the metric to be Schwarzschild for all spacetime dimensions $d$ and Lovelock polynomial order $N$. In the following section the metric ansatz of Finch Skea is analysed for the $N=2$ pure Gauss-Bonnet case and the exact models for $d=5, 6$  are obtained explicitly. We then demonstrate similarity between $d=3, 4$ Einstein  with $d=5, 6$ pure GB solutions. We conclude with a discussion of our results.

\section{Pure Lovelock gravity}

The Lovelock polynomial action \cite{lov} is given by the Lagrangian
\begin{equation}
\mathcal{L} = \sum ^N_{N=0} \alpha_N \mathcal{R}^{(N)}  \label{5}
\end{equation}
where
\begin{equation}
 \mathcal{R}^{(N)} = \frac{1}{2^N} \delta^{\mu_1 \nu_1 ...\mu_N \nu_N}_{\alpha_1 \beta_1 ... \alpha_N \beta_N} \Pi^N_{r=1} R^{\alpha_r \beta_r}_{\mu_r \nu_r}
\end{equation}
 and $\mathcal{R}^{\alpha \beta}_{\mu \nu}$
 is the $N$th order Lovelock analogue of the Riemann tensor as defined in Ref. \cite{bianchi}.  Also $ \delta^{\mu_1 \nu_1 ...\mu_N \nu_N}_{\alpha_1 \beta_1 ... \alpha_N \beta_N} = \frac{1}{N!} \delta^{\mu_1}_{\left[\alpha_1\right.} \delta^{\nu_1}_{\beta_1} ... \delta^{\mu_N}_{\alpha_N} \delta^{\nu_N}_{\left.\beta_N \right]}$ is the required Kronecker delta. \\

On variation of the action including the matter Lagrangian with respect to the metric,  we get the equation of motion \cite{bianchi}
\begin{equation}
 \sum ^N_{N=0} \alpha_N \mathcal{G}^{(N)}_{AB} = \sum ^N_{N=0} \alpha_N \left( N\left(\mathcal{R}^{(N)}_{AB} - \frac{1}{2}\mathcal{R}^{(N)}g_{AB}\right)\right) =  T_{AB} \label{7}
\end{equation}
where $ \mathcal{R}^{(N)}_{AB} = g^{CD}\mathcal{R}^{(N)}_{ACBD},  \mathcal{R}^{(N)}=g^{AB}\mathcal{R}^{(N)}_{AB}$, and $T_{AB}$ is the energy momentum tensor. Note that the trace of the Bianchi derivative of $N$th order Riemann tensor, $\mathcal{R}^{\alpha \beta}_{\mu \nu}$, yields the corresponding Einstein tensor, $\mathcal{G}^{(N)}_{AB}$. This is the gravitational equation in Lovelock gravity which corresponds to the cosmological constant for $N=0$, to Einstein's equation for $N=1$ and to the Gauss-Bonnet equation for $N=2$, and so on.  \\

In particular the Einstein-Gauss-Bonnet equation is given by
\begin{equation}
G^{A}_{B} + \alpha H^{A}_{B} = T^{A}_{B}  \label{6}
\end{equation}
 where $\mathcal{G}^{(2)}_{AB} = H_{AB}$ and
\begin{equation}
H_{AB} = 2\left(R R_{AB} - 2R_{AC}R^C_B - 2R^{CD}R_{ACBD} + R^{CDE}_{A} R_{BCDE} \right) - \frac{1}{2} g_{AB} \mathcal{R}^{(2)}.
\end{equation}

Henceforth we would specialize to the pure Lovelock equation involving a single, $\mathcal{G}^{(N)}_{AB}$, on the left corresponding to the $N$th term without sum over the lower orders. The equation is then given by

\begin{equation}
 \mathcal{G}^{(N)}_{AB} =  \left( N\left(\mathcal{R}^{(N)}_{AB} - \frac{1}{2}\mathcal{R}^{(N)}g_{AB}\right)\right) =  T_{AB} . \label{7}
\end{equation}

\section{Pure Lovelock star solutions}

The general static $d-$dimensional spherically symmetric metric is taken  to be
\be
ds^2 = - e^{\nu} dt^2 + e^{\lambda} dr^2 + r^2 d\Omega^2_{d-2} \label{8a1}
\ee
where $d\Omega^2_{n-2}$ is the metric on a unit $(d-2)$-sphere and where $\nu = \nu(r)$ and $\lambda = \lambda (r)$ are the metric potentials.
  The energy--momentum tensor for the
comoving fluid
velocity vector $u^a=e^{-\nu/2} \delta^{a}_{0}$
has the form
$
T^a_{b}= \mbox{diag} \left( -\rho,\, p_r ,\,  p_{\theta} ,\,
p_{\phi},\, ...\right)$
for a neutral perfect fluid. Note that in view of spherical symmetry we have for all the $(d-2)$ angular coordinates $p_{\theta} = p_{\phi} = ....$.
The conservation laws   $T^{ab}_{}{}_{;b} = 0 $  result in the single equation
\be
\frac{1}{2}\left( p_r + \rho\right)\nu' + p'_r + \frac{(d-2)}{r}\left(p_{r} - p_{\theta}\right) = 0 \label{8a11}
\ee
where primes denote differentiation with respect to $r$. The expressions for density, radial and transverse pressure obtained from the pure Lovelock equation are given by
\begin{eqnarray}
\rho &=&\frac{(d-2) e^{-\lambda} \left(1-e^{-\lambda}\right)^{N-1}\left(rN\lambda' + (d-2N-1)(e^{\lambda} -1)\right)}{2r^{2N}} \label{9a} \\
p_r &=&\frac{(d-2) e^{-\lambda} \left(1-e^{-\lambda}\right)^{N-1}\left(rN\nu' - (d-2N-1)(e^{\lambda} -1)\right)}{2r^{2N}}  \label{9b} \\
p_{\theta} &=& \frac{1}{4}  r^{-2 N} \left(e^{\lambda}\right)^{-N} \left(e^{\lambda}-1\right)^{N-2} \left[-N r \lambda' \left\{2 (d-2 N-1) \left(e^{\lambda}-1\right)  \right. \right. \nonumber \\
&& \left. \left.
+r \nu' \left( e^{\lambda}-2 N +1\right)\right\} + \left(e^{\lambda}-1\right) \left\{-2 (d-2 N-1) (d-2 N-2) \left(e^{\lambda}-1\right) \right. \right. \nonumber \\
&& \left. \left.
+2N  ( d-2N-1) r\nu' +Nr^2 \nu'^2 +2N r^2 \nu''\right\}\right] \label{9c}
\end{eqnarray}
where the Lovelock coupling parameter has been set to unity.

The equation of pressure isotropy $p_r = p_{\theta}$ reads as
\beq
 r\lambda' \left\{2 (d-2 N-1)  \left(e^{\lambda}-1\right)+r\nu' \left( e^{\lambda}-2 N +1\right)\right\} && \nonumber \\
  -\left(e^{\lambda}-1\right) \left\{4 (d-2 N-1) \left(e^{\lambda}-1\right)+r \left(\nu' \left(-4 N+r \nu'+2\right)+2 r \nu''\right)\right\} &=& 0 \label{9d}
\eeq

Invoking change of coordinates as $x=Cr^2$, $e^{-\lambda}= Z(x)$ and $e^{\nu} = y^2(x)$ we write Eqns (\ref{9a}) to (\ref{9c}) as

\begin{eqnarray}
\rho &=&\frac{C^{N} (d - 2) (1 - Z)^{N - 1} \left[ (d - 2 N - 1) (1 - Z) -2 N x \dot{Z} \right]  }{2 x^{N}}   \label{10a} \\
p_r &=&\frac{C^{N} (d - 2)  (1 - Z)^{N - 1} \left[ 4 N x Z \dot{y} - (d - 2 N - 1) (1 - Z) y \right]  }{2 x^{N} y}  \label{10b} \\
p_{\theta} &=& \frac{C^{N}  (1 - Z)^{N - 2} }{2 x^{N} y} \left[ 8 N x^{2} Z (1 - Z) \ddot{y} \right. \nonumber \\
&+& \left.  4 N x \left( x (1 + (1 - 2 N) Z) \dot{Z} + (d - 2 N) Z ( 1 - Z) \right) \dot{y} \right. \nonumber \\
&+& \left. (d - 2 N - 1) (1 - Z) \left( 2 N x \dot{Z} - (d - 2 N -2) (1 - Z) \right) y \right] \label{10c}
\end{eqnarray}

while the pressure isotropy assumes the form
\beq
4x^2Z(1-Z)\ddot{y} + \left[ 4(1-N)x Z (1-Z) +2x^2\left(1-(2N-1)Z\right)\dot{Z}\right]\dot{y} && \nonumber  \\ + (d-2N-1)(1-Z)\left(\dot{Z}x -Z + 1\right) y &=& 0 \label{9l}
\eeq
Setting $N=1$ yields the standard $d-$ dimensional equation for Einstein gravity
\be
4x^2Z\ddot{y} +  2x^2\dot{Z}\dot{y}  + (d-3)\left(\dot{Z}x -Z + 1\right) y = 0 \label{9m}
\ee
as discussed in \cite{chil-hans2}. For the pure Gauss--Bonnet, $N=2$ case Eqn (\ref{9l}) reads as
\beq
4x^2Z(1-Z)\ddot{y} + \left[ -4xZ(1-Z) +2x^2(\left(1-3Z\right)\dot{Z}\right]\dot{y} && \nonumber  \\ + (d-5)(1-Z)\left(\dot{Z}x -Z + 1\right) y &=& 0 \label{9n}
\eeq \\

This is the only equation to be solved and it involves two unknown functions $\nu$ and $\lambda$. It cannot be solved without an additional prescription, either on the metric potential or a relation between density and pressure -- an equation of state, or fall off behavior for density. For constant density, there is the unique Schwarzschild interior solution \cite{dkm} in all dimensions $d \geq 4$ in Einstein as well as in Lovelock theory. However for Lovelock gravity, the solution exists for dimensions $d\geq2N+2$  while there cannot exist any bound distribution in critical odd dimensions $d=2N+1$. For an isothermal distribution characterized by $\rho\sim 1/r^{d-2}$, it has recently been shown \cite{isothermal} that there is a universal solution like the Schwarzschild interior solution for uniform density for  pure Lovelock gravity in all dimensions $d\geq 2N+2$. \\

\subsection{The case $d=2N+1$}

In the critical odd $d=2N+1$, gravity is kinematic for pure Lovelock gravity \cite{bianchi, cam-d}, which means vacuum is Lovelock flat. Hence it cannot be an exterior to any matter distribution. This fact reflects the absence of a finite non-zero radius where $p=0$, defining the boundary for fluid distribution. In the odd critical dimensions $d=2N+1$, the fluid distribution has to be unbounded -- cosmological. We shall demonstrate this feature in the following examples.

Setting $d = 2N + 1$, the equations of motion reduce to
\begin{eqnarray}
\rho &=& -C^{N} (2N-1)N (1 - Z)^{N - 1}    x^{1-N} \dot{Z}     \label{11e} \\ \n \\
p_r &=&\frac{2C^{N} (2N-1)N  (1 - Z)^{N - 1}  x^{1-N} Z \dot{y}    }{  y}  \label{11f} \\ \n \\
0 &=& 2xZ(1-Z)\ddot{y} + \left[ 2(1-N) Z (1-Z) +x\left(1-(2N-1)Z\right)\dot{Z}\right]\dot{y}  \label{11a}
\end{eqnarray}
A first integral of equation (\ref{11a}) may be obtained in the form
\be
\dot{y}=\frac{K_1 x^{N-1} (1-Z)^{1-N}}{Z^{\frac{1}{2}}} \label{11g}
\ee
Observe that for the pressure to vanish at any non-zero finite radius in order to admit a bounded distribution, this amounts to $Z = 1$, $Z = 0$ or $\dot{y} = 0$ from equation (\ref{11f}). The first option $Z=1$ leads to zero density from Eqn (\ref{11e}) and we are back to vacuum. Putting $Z = e^{-\lambda} = 0$ is obviously inadmissable. Finally $\dot{y} = 0$ does cause Eqn (\ref{11a}) to be identically satisfied, however the constant $\nu$ is not compatible with matching to the exterior vacuum solution. Additionally $\dot{y} =0$ forces $Z=1$ in view of (\ref{11g}).  Therefore we must conclude that in general no bounded compact sphere can exist for all $N$ in the critical odd dimensions $d = 2N +1$.

Let us in particular consider the two examples of incompressible fluid sphere and Finch-Skea model \cite{Finch-Skea} in the critical odd dimension.

\begin{itemize}
\item {Incompressible fluid sphere}

Consider the special case of  an incompressible fluid sphere $\rho = $ constant $=\rho_0$. Equation (\ref{11e}) integrates as
\be
Z= 1-\left(ax^N +c_1\right)^{\frac{1}{N}} \label{11a1}
\ee
 where $c_1$ is an integration constant and we have redefined $a = \frac{\rho_0}{c^N N (2N-1)}$. Plugging this in Eqn (\ref{11a}) yields
\be
y= e^{\nu/2} = C_2-\frac{2 C_1}{a} \sqrt{Z} = C_2-\frac{2 C_1}{a} e^{-\lambda/2} \label{11a2}
\ee
where $C_1$ and $C_2$ are integration constants. The pressure has the form
\be
p=\frac{2 C_1 \rho_0 e^{-\lambda/2}}{ C_2 - 2 C_1 \sqrt{1-\left(a x^N+ c_1\right)^{1/N}}}. \label{11a3}
\ee
Clearly $p=0$ requires $e^{-\lambda}=0$ which is unacceptable, and hence there exists no boundary to the distribution.

\item{Finch--Skea model}

For the Finch--Skea ansatz \cite{Finch-Skea} we set $Z=\frac{1}{1+x}$, then Eqn (\ref{11a}) integrates to give
\be
y=\frac{2 c_1 (x+1)^{N+\frac{1}{2}}}{2 N+1}+c_2  \label{11b}
\ee
and we have the density and pressure given by
\beq
\rho &=&  \frac{N (2 N-1) C^N}{ (x+1)^{N+1}}  \label{11f} \\ \n \\
p&=& \frac{2 c_1 N \left(4 N^2-1\right) C^N}{\sqrt{x+1} \left(2 c_1 (x+1)^{N+\frac{1}{2}}+2 c_2(2N+1)\right)}  \label{11g}
\eeq

Here again the pressure cannot vanish without collapsing to vacuum and hence there cannot exist any boundary to the distribution. This is the general property in all odd critical dimensions $d=2N+1$. It has always to be an
unbounded cosmological distribution.\\
\end{itemize}

\section{Uniform density Schwarzschild solution}

It has been shown that a uniform density sphere is always described by the Schwarzschild interior solution \cite{dkm} for Einstein as well as for Lovelock theory in any dimension $d \geq 4$. However for pure Lovelock gravity as argued above, it is Schwarzschild for dimensions $d\geq 2N+2$ dimensions but not for the critical odd dimensions $d=2N+1$. We shall obtain the general solution for constant density for an arbitrary $N$. \\

Observe that Eqn (\ref{9a}) can be rearranged into the form
\be
\rho = \frac{(d-2)\left[ r^{d-2N-1} \left( 1 - e^{-\lambda}\right)^N\right]'}{2r^{d-2}} \label{9e}
\ee
which for a constant density integrates to give
\be
e^{-\lambda} = 1-\left(\frac{2\rho_0}{(d-2)(d-1)} r^{2N} + K_2r^{2N+1-d}\right)^{1/N} \label{9o}
\ee
where $K_2$ is an integration constant. Since $d\geq 2N+2$, $K_2=0$ for avoiding the singularity at the center. Thus we have
\be
e^{-\lambda} = 1-\frac{2\rho_0}{(d-2)(d-1)} r^{2}  \label{9q}
  \ee
which is clearly of the Schwarzschild form. Then the isotropy equation integrates to give the general solution
\be
e^{\nu/2} = c_2 + c_1 e^{-\lambda/2} \label{9m}
\ee
where $c_1$ and $c_2$ are constants of integration. This is the Schwarzschild solution and it is universal for all dimensions $ d \geq 2N+2$. \\

Alternatively if we assume $Z=1+x$, the isotropy equation (\ref{9d}) becomes
 \[
2(1+x) \ddot{y} + \dot{y} = 0
 \]
and integration  gives
\be
y(x) = c_1 \sqrt{1+x} +c_2 \label{1012}
\ee
equivalent to Eqn (\ref{9m}). This shows that the ansatz $Z=1+x$ is equivalent to the density being constant as it gives the Schwarzschild solution independent of spacetime dimension or order of the Lovelock polynomial \cite{dkm}.

For this model the density and pressure are given by
\beq
\rho &=& \frac{C^N (d - 1)(d - 2)}{2}, \label{1013} \\
  p &=& \frac{C^N (d - 2) \left[ c_1 (1 - d) \sqrt{1 + x} - c_2 (d - 2 N - 1) \right]}{2 \left(  c_1 \sqrt{1+x} + c_2 \right)}. \\
\eeq

Clearly pressure vanishes at some finite $r$ defining boundary of the distribution, and it could then be matched to pure Lovelock analogue \cite{probes3} of the Schwarzschild exterior solution determining the constants $c_1$ and $c_2$. Since the density is constant the sound speed would be infinite which despite being unphysical  it  defines the Buchdahl compactness limit  \cite{comp-limit}.

\section{Finch-Skea potential ${\displaystyle Z=\frac{1}{1+x}}$}

It is interesting to investigate other choices of potentials that yield completely solvable models. In particular the Finch--Skea model \cite{Finch-Skea} is a physically viable star interior model as argued in \cite{wal}. Here one of the metric potentials is prescribed as $Z = \frac{1}{1 + x}$, then from Eqn (\ref{9l}) we obtain
\begin{equation}
4 (1 + x) \ddot{y} - 2 (2 N - 1) \dot{y} + (d - 2 N - 1) y = 0. \label{9p}
\end{equation}
Let $ V = 1 + x $, so that Eqn (\ref{9p}) becomes
\begin{equation}
4 V \frac{d^2 y}{d V^2} - 2 (2 N - 1) \frac{d y}{d V} + (d - 2 N - 1) y = 0. \label{9q}
\end{equation}
Assuming $ y(V) = u(V) V^m $, we transform equation (\ref{9q}) to
\begin{equation}
4 V^2 \frac{d^2 u}{d V^2} + \left[ 8 m - 2 (2 N - 1) \right] V \frac{d u}{d V} + \left[ 4 m^2 - 2 m (1 + 4 N) + (d - 2 N - 1) V \right] u = 0. \label{9r}
\end{equation}
We now let $ z = V^{\beta} $, which takes Eqn (\ref{9r}) to
\beq
 4 \beta ^{2} Z^2 \frac{d^2 u}{d z^2} + \left[ 4 \beta (\beta - 1) \beta (8 m - 2 (2 N - 1)) \right] z \frac{d u}{d z}
&& \nonumber \\
 + \left[ 4 m^2 - 2 m (1 + 4 N) + (d - 2 N - 1) z^{\frac{1}{\beta}} \right] u = 0. \label{9s}
\eeq
Setting $ m = 1 $ and $ \beta = \frac{1}{2}$, it reduces to
\begin{equation}
z^2 \frac{d^2 u}{d z^2} + 2 (2 - N) z \frac{d u}{d z} + \left[ (d - 2 N - 1) z^2 + 2 (1 - 4 N) \right] u = 0 \label{9t}
\end{equation}
This can further be transformed to a differential equation of the Bessel type and consequently it yields the general solution for Eqn (\ref{9p})  given by
\be
y = (1+x)^{(2N+1)/4} \left(c_1 J_{-N-\frac{1}{2}} \left(\sqrt{(d-2N-1)(1+x)}\right) + c_2 Y_{-N-\frac{1}{2}} \left(\sqrt{(d-2N-1)(1+x)}\right) \right) \label{9u}
\ee
where $J$ and $Y$ are Bessel functions of the first and second kind respectively. Since the Bessel functions are of half-integer order  they can be written in terms of elementary functions through the formulae
\[
J_{-N-\frac{1}{2}}(w) = \sqrt{\frac{2}{\pi}}w^{N+\frac{1}{2}} \left(\frac{1}{w}\frac{d}{dw}\right)^N \frac{\cos w}{w}  \hspace{20mm}  Y_{-N-\frac{1}{2}}(w) = \sqrt{\frac{2}{\pi}}w^{N+\frac{1}{2}} \left(\frac{1}{w}\frac{d}{dw}\right)^N \frac{\sin w}{w}
\]
for all orders of degree $N$ of the Lovelock polynomial. Thus we have completely solved the isotropy equation for general $N$ and $d$ for the Finch--Skea ansatz $Z=\frac{1}{1+x}$ and the solutions may be resolved in terms of elementary functions.

\subsection{Einstein case $N=1$}

Setting $N=1$ in Eqn (\ref{9u}) generates the same solution as reported in \cite{chil-hans} for the higher dimensional Finch--Skea solution. It is given by the metric,
\begin{eqnarray}
d s^{2} & = & \left( d - 3 \right) ^{-3} \left[ \left( C_{1} - C_{2} v \right) \sin v - \left( C_{2} + C_{1} v \right) \cos v \right] ^{4} d t ^{2} \nonumber \\
        & \quad & - \left[ \left( 1 + C r^{2} \right) ^{2} \right] ^{-1} d r^{2} - r^{2} d \Omega _{d - 2} ^{2} \label{13},
\end{eqnarray}
where $ v = \sqrt{( d - 3) ( 1 + C r^{2})} $ and $C_1$ and $C_2$ are constants.

Note that the case $d=3$ is excluded from this class of solutions. It corresponds to $d=2N+1$ when $N=1$ and it is a case that has been treated separately and note that the coefficient of $y$ vanishes in Eqn (16). Three dimensional solutions can be found, however they do not emerge from this treatment. We must set $d=3$ in Eqn (\ref{9m}) to generate the appropriate solution but this is not of interest to our present work.

The dynamical quantities energy density and pressure are respectively given by
\begin{equation}
\rho =  \frac{ C (d - 2)(d-3)^2 \left[ v^{2} + 2 \right] }{2 v^4}, \label{14}
\end{equation}
and
\begin{equation}
p = \frac{C (d - 2) (d - 3)^2}{ 2 v^2} \left[ \frac{ \left( \zeta + v \right) \tan v + \zeta v - 1 }{ \left( \zeta - v \right) \tan v -\zeta v - 1 } \right], \label{15}
\end{equation}
where $ \zeta = \frac{C_{1}}{C_{2}} $.

\subsection{Pure Gauss--Bonnet case $N=2$}

As discussed earlier the isotropy equation (\ref{9l}) in the critical odd dimensions $d=2N+1$ takes an ostensibly first order form because the coefficient of $y$ vanishes. For pure GB, we shall consider the critical odd and even dimensions $d=5, 6$.\\

\begin{itemize}
\item{\bf $N=2$, $d=5$}

In this case we have for equation (\ref{9l}),
\begin{equation}
2xZ(1-Z)\ddot{y} + \left(2Z(Z-1) + (1-3Z)x\dot{Z} \right)\dot{y} = 0.  \label{100d}
\end{equation}
for $d = 5$ and $N=2$.
It is a second order differential equation in $y$, however it could as well be written as a first order ordinary equation in $Z$, and thereby we have
 \begin{equation}
 y=C_1 \int{xe^{\int \frac{(3Z-1)\dot{Z}}{2Z(1-Z)}dx}dx} + C_2 \label{1111}
 \end{equation}
 where $C_1$ and $C_2$ are constants of integration. The functional forms for $Z(x)$ may be selected {\it a priori} so as to allow for the complete integration of the equation. Note that $Z = const.\neq 1$  gives $y=C_1x^2 + C_2$. However it is not tenable as the density vanishes in Eqn (13) \cite{isothermal}. The Finch--Skea ansatz $Z=\frac{1}{1+x}$ results in the solution
\be
e^{\nu} = c_1(1+Cr^2)^{\frac{5}{2}} + c_2 \label{1201}
\ee
where we have reverted to the original coordinates via $x=Cr^2$ for greater transparency.
The energy density is now given by
\be
\rho = \frac{12  C^3}{(1+Cr^2)^3} \label{1202}
\ee
while the pressure has the form
\be
p= \frac{60   C^3}{ (1+Cr^2)^{\frac{1}{2}} \left(\left( 1+C r^2\right)^{\frac{5}{2}}+ \kappa \right)}\label{1203}
\ee
where we have put $\kappa = \frac{C_2}{C_1}$. Observe that it is impossible for the pressure to vanish for any finite radius, therefore the solution cannot describe a finite object but, instead, a cosmological fluid.
The quantity
\be
\frac{dp}{d\rho} = \frac{5  \left(C r^2+1\right)^{\frac{5}{2}} \left(6  \left(C r^2+1\right)^{\frac{5}{2}}+ \kappa \right)}{6 \left( \left(C r^2+1\right)^{\frac{5}{2}}+ \kappa\right)^2} \label{1204}
\ee
represents the sound-speed in general relativity and we consider its value in the present context as well.
Interestingly by solving for $r$ in Eqn (\ref{1202}) and substituting in Eqn (\ref{1203}) we obtain a barotropic equation of state $p=p(\rho)$ which has the form
\be
p= \frac{60 C^3 \rho^{\frac{11}{6}}}{12 C^3 \left( (12 C^3)^{\frac{5}{6}} + \kappa \rho^{\frac{5}{6}}\right)} \label{1205}
\ee
This is usually a desirable feature in perfect fluid configurations but it is not often obtained for exact solutions of the standard Einstein field equations. \\

We now compare the 5-dimensional solution given above to its 3-dimensional counterpart in Einstein gravity. The
Einstein's field equations read as
\beq
\rho &=& C\dot{Z} \label{1402a} \\ \n \\
p_r &=& \frac{-2CZ\dot{y}}{y} \label{1402b} \\ \n \\
p_{\theta} &=& \frac{-4CxZ\ddot{y} - 2CZ\dot{y} - 2Cx\dot{Z} \dot{y}}{y} \label{1403c}
\eeq
in our scheme. The equation of pressure isotropy assumes the simple form $2Z\ddot{y} + \dot{Z} \dot{y} =0 $  which integrates as
\be
y(x)=c_1 \int Z^{-\frac{1}{2}} dx + c_2.
\ee

As was the case for $5$-dimensional GB, here too  the case $Z= const.$ is not tenable. The Finch--Skea ansatz $Z=\frac{1}{1+x}$  results in the solution
\be
y=c_1 (1+ Cr^2)^{\frac{3}{2}} + c_2  \label{1404}
\ee
with the density and pressure  given by
\be
\rho = \frac{C}{(1+Cr^2)^2} \hspace{1cm} p =  \frac{3C}{(1+Cr^2)^{\frac{1}{2}} \left((1+Cr^2)^{\frac{3}{2}}+\kappa\right)} \label{1407b}
\ee
respectively (See Ref. \cite{Banerjee} for an independent derivation). \\

From Eqns (\ref{1201}), (\ref{1202}) and (\ref{1404}) the similarity between the two solutions corresponding to critical odd dimensions $d=2N+1$ for $N=1, 2$ is evident, and there exists no zero pressure surface. A comparison of the density and pressure functions for $d = 3$ and $5$ is shown in table 1.  \\

\begin{table}
\caption{\bf{Energy density and pressure for $d=2N+1$ ( $N=1, 2$) }}

\medskip

\begin{tabular}{|c|c|c|}
  \hline
  & $d =3$ Einstein & $d=5$ GB \\
  \hline
 $\rho$ &$ \frac{C}{(1+Cr^2)^2}$ & $\frac{12C}{(1+Cr^2)^3}$ \\
 \hline
 $ p$ & $\frac{60C}{(1+Cr^2)^{\frac{1}{2}} \left((1+Cr^2)^{\frac{3}{2}} + \kappa\right)}$ & $\frac{3C}{(1+Cr^2)^{\frac{1}{2}} \left((1+Cr^2)^{\frac{5}{2}} + \kappa\right)}$ \\
  \hline
\end{tabular}
\end{table}


\item{\bf $N=2$, $d=6$}

The equation of pressure isotropy (\ref{9l}) assumes the form
\begin{eqnarray}
4 x^2  Z   ( 1-Z)   \ddot{y}  +  2 x\left(x\dot{Z} (1-3Z) - 2Z(1-Z)\right)   \dot{y} \nonumber \\ + (1-Z) (x\dot{Z} -Z + 1)  y &=& 0 .\label{67c}
\end{eqnarray}
 With the transformation $Z = \frac{1}{1+x}$, Eqn (\ref{67c}) reduces to
\be
4(1+x)\ddot{y} - 6\dot{y} +y=0 \label{60j}
\ee
Redefining $1+x = X$, it becomes
\be
4X\frac{d^2 y}{dX^2} - 6\frac{dy}{dX} + y=0 \label{60k}
\ee
and making the substitution $y(X)=u(X)X^m$ for a new function $u$ and constant $m$, it further transforms to
\be
4X^2 \ddot{u} + (8m - 6 )X\dot{u} + (4m^2 - 10m + X)u=0 \label{60l}
\ee
where dots are now derivatives with respect to $X$. Invoking a transformation $v=X^{\gamma}$, we have
\be
4\gamma^2 X^{2\gamma} \frac{d^2 u}{dv^2} + (4\gamma (\gamma - 1) + (8m - 6)\gamma)X^{\gamma} \frac{du}{dv} + (4m^2 -10m + X)u=0. \label{60m}
\ee
The choice $m=1$ and $\gamma = \frac{1}{2}$ results in the form
\be
v^2\frac{d^2 u}{dv^2} + (v^2 - 6)u=0 \label{60n}
\ee
which is solved by the spherical Bessel functions $J_n$ and $Y_n$ when $n=2$. Interestingly for $n=2$ we are able to find solutions in terms of simple elementary functions given by
\be
u(v) = a \left(\frac{3 \sin v - v^2 \sin v -3v\cos v}{v^2}\right) + b \left( \frac{v^2 \cos v -3\cos v -3v\sin v}{v^2}\right) \label{60o}
\ee
where $a = \sqrt{\frac{2}{\pi}} c_1$ and $b = \sqrt{\frac{2}{\pi}} c_2$ are arbitrary constants and it  must be recalled that $v= X^{\frac{1}{2}} = \sqrt{1+x} = \sqrt{1+Cr^2}$ in the original canonical variables.
Now the gravitational potentials and dynamical quantities have the form
\beq
e^{\lambda} &=& v =  \sqrt{1+Cr^2} \label{70a} \\ \n \\
e^{\nu} &=&  \left(a (3 - v^2) - 3 b v\right)\sin v + \left( b (v^2 - 3) - 3 a v \right) \cos v   \label{70b} \\ \n \\
\rho&=& \frac{12   (v^2+4)}{v^6} \label{70c} \\ \n \\
p&=& \frac{2  \left[ (\zeta v - v^2 - 1) \tan v + \zeta (v^2 + 1) + v \right] }{
 v^6 \left[ \left( v^2 - 3 + 3 \zeta v \right) \tan v - \zeta \left( v^2 - 3 \right) + 3 v
 \right]} \n \\  \label{70d}
\eeq
where we have defined $\zeta = \frac{a}{b}$ and put $C=1$.

Following the standard approach in general relativity we examine the quantity
\beq
\frac{dp}{d\rho} &=& - \frac{v^2 \left[ 2 (v^4 - 3) - \zeta v (v^2 - 18 + 12 \zeta v) \right] \tan ^2 v}{ 2 (v^2 + 6) \left[ \left( v^2 - 3 + 3 \zeta v \right) \tan v - \zeta \left( v^2 - 3 \right) + 3 v
 \right]^2} \nonumber \\
                 & \quad & + \frac{v^2 \left[ 2 v (v^3 - 3) + \zeta (4 v^3 + 15 v^2 + 12 v + 6 ) \right] \tan v}{ 2 (v^2 + 6) \left[ \left( v^2 - 3 + 3 \zeta v \right) \tan v - \zeta \left( v^2 - 3 \right) + 3 v
 \right]^2} \nonumber \\
                 & \quad & - \frac{  \zeta ^2 v^2 \left[ 2 v^4 + 6 v^3 - 3 v^2 - 9 v - 6 \right] \tan v}{ 2 (v^2 + 6) \left[ \left( v^2 - 3 + 3 \zeta v \right) \tan v - \zeta \left( v^2 - 3 \right) + 3 v
 \right]^2} \nonumber \\
                 & \quad & + \frac{v^2 \left[ 3 v (v^2 - 1) - \zeta \left( v^4 - 7 v^2 + 3 + \zeta v (v^2 - 3) \right) \right] \sec ^2 v}{ 2 (v^2 + 6) \left[ \left( v^2 - 3 + 3 \zeta v \right) \tan v - \zeta \left( v^2 - 3 \right) + 3 v
 \right]^2} \nonumber \\
                  & \quad & + \frac{v^2 \left[ 18 v^2 - \zeta \left( 3 v (v^2 - 5) - 2 \zeta (5 v^2 + 3) \right) \right]}{ 2 (v^2 + 6) \left[ \left( v^2 - 3 + 3 \zeta v \right) \tan v - \zeta \left( v^2 - 3 \right) + 3 v
 \right]^2} \label{801}
\eeq
which  is constrained to be less than unity for a subluminal speed of sound in general relativity. We contrast the relativistic value of this index with its analogue in pure Lovelock theory.

 The behaviour of the dynamical quantities at the stellar centre ($r=0$, that is $v=1$) will assist in establishing bounds on $\beta$. The positivity of energy density and pressure requires
 \beq
 \rho_0 &=& 60 C > 0 \label{80a} \\ \n \\
 p_0 &=& \frac{-11 \zeta+(-2 \zeta-11) \tan (1)+2}{3 \zeta+(3-2 \zeta) \tan (1)+2} >0 \label{80b}
 \eeq
 which immediately forces $ C > 0$ and so we set $C = 1$ henceforth.  Requiring the distribution to have positive energy density and pressure at the centre as well as the satisfying of causality leads to the condition $ \zeta = \frac{c_1}{c_2}  < 0.017208$ or $\zeta > 0.594482$. For the purpose of generating plots of the dynamical quantities we select the value $\zeta = 1$.

\begin{figure}
  \includegraphics[width=0.8\textwidth]{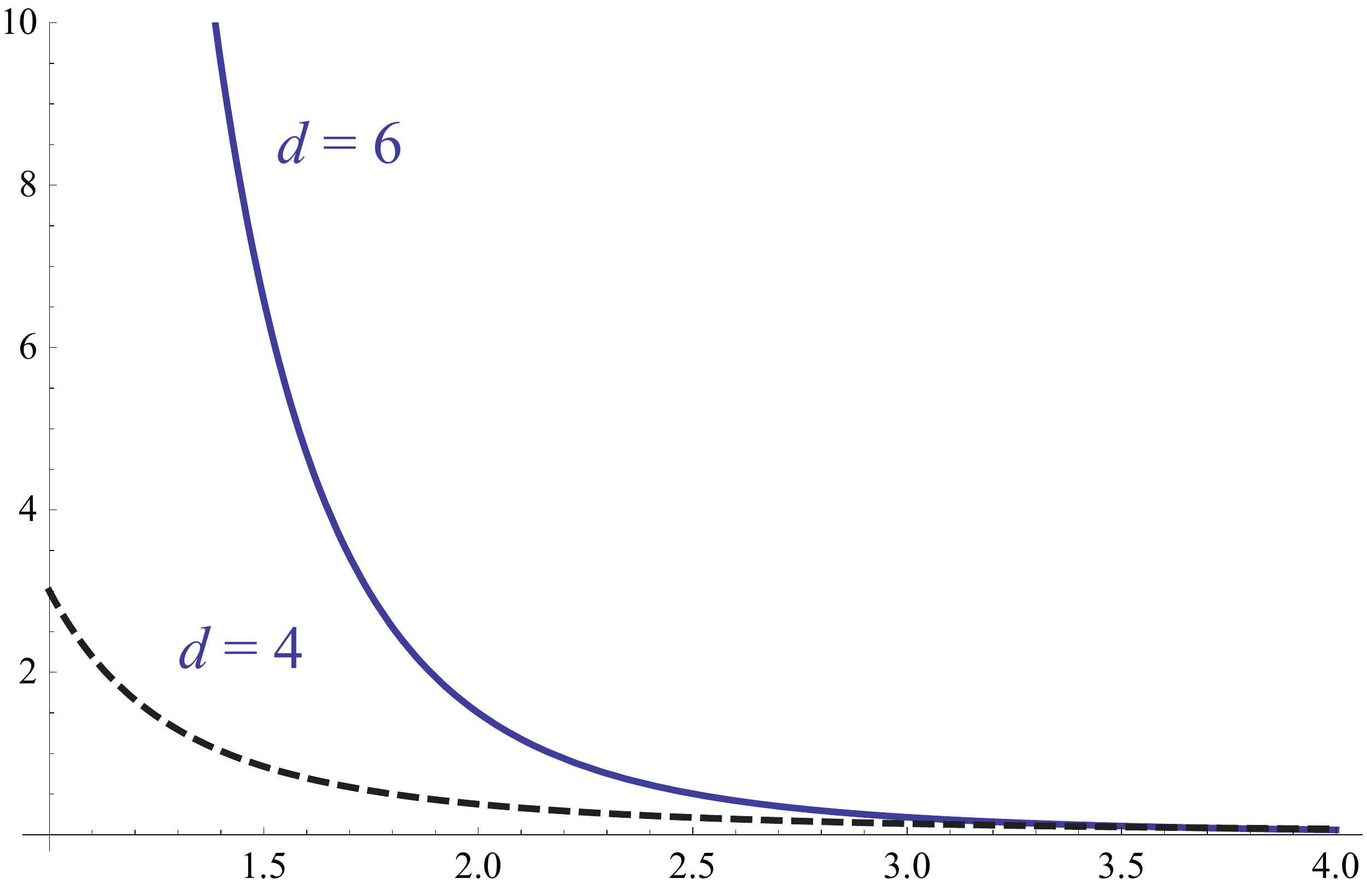}\\
  \caption{Plot of energy density versus radius for $\zeta = 1$ in 4 and 6D GB}\label{Fig. 4}
\end{figure}
\begin{figure}
  \includegraphics[width=0.8\textwidth]{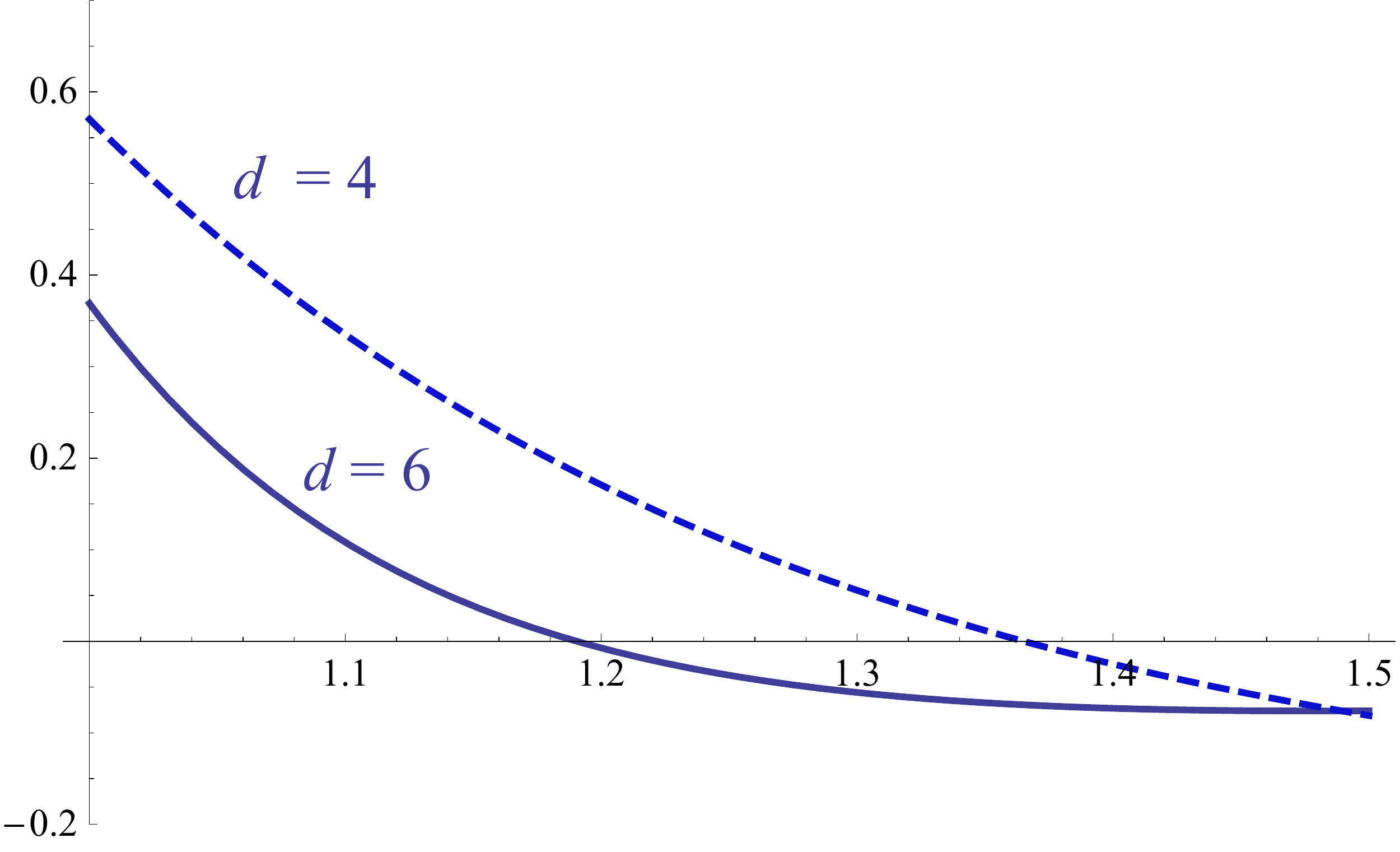}\\
  \caption{Plot of pressure versus radius for $\zeta = 1$ in 4 and 6D GB}\label{Fig. 5}
\end{figure}

\begin{figure}
  \includegraphics[width=0.8\textwidth]{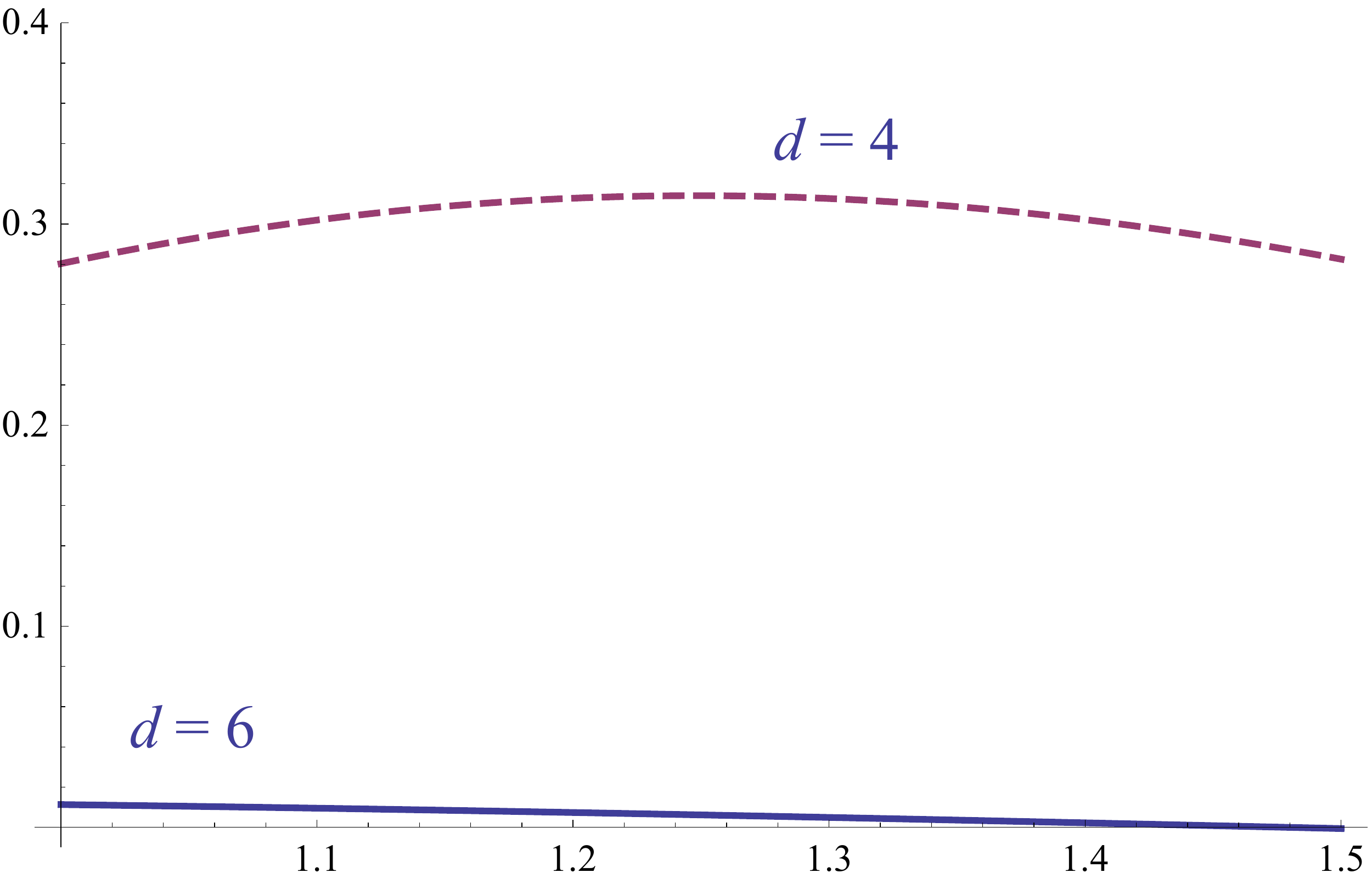}\\
  \caption{Plot of sound--speed index versus radius for $\zeta = 1$ in 4 and 6D GB}\label{Fig. 6}
\end{figure}

From the plots in Figs 1 -- 3, it can be seen that there exists a pressure zero surface at approximately $v= 1.2$ in $d=6$ and $v = 1.4$ in $d=4$ which identifies the boundary of the distribution.  Within this radius, it is clear that  pressure and energy density are positive definite. Fig. 3 demonstrates that the fluid is causal. Note that the dynamical quantities are dependent on periodic functions. It is therefore necessary to identify the first zero of the numerator of the pressure function in order to fix the boundary of the distribution.

In the four dimensional Einstein scenario the Finch and Skea solution \cite{Finch-Skea}   is
given by
\beq
e^\lambda&=& v \label{574a}\\ \n \\
e^\nu&=& (c_1 +c_2 v)\sin v  + (\:c_2-c_1
v\:)\:\cos v\:  \label{574b} \\ \n \\
\rho&=&\frac{C(v^2 +2)}{v^4}\label{574c}\\ \n \\
p&=& \frac{C\left( \left( \zeta v +1 \right)+\left(
\zeta-
v \right) \tan v\right)}{v^2\left( \left( 1- \zeta
v\right)
+ \left( \zeta+ v \right) \tan v\right)}\label{574d}
\eeq
From Eqn (\ref{574c}) it is clear that $C>0$ for a positive energy density. Positivity of the central pressure ($p_0 >0$) then yields $\zeta < -4.58804$ or $\zeta > 0.217958$ as found by Finch and Skea. The causality criterion $0<\left(\frac{dp}{d\rho}\right)_0 < 1$, imported from general relativity,  results in the restriction $-2.6141 < \zeta < -1.68214$ or $-0.642093 < \zeta < 6.40698$. Harmonising all these conditions requires that $0.217958 < \zeta < 6.40698$. We select the value $\zeta = 1$ in order to plot the dynamical quantities.


The critical even $d=2N+2$-dimensional solutions for $N=2$ GB and $N=1$ Einstein are given by Eqns (\ref{70a}) -- (\ref{70d}) and (\ref{574a}) -- (\ref{574d}) exhibiting the similarity between the two (See Table II).  The plots of the 6-dimensional pure Lovelock and 4-dimensional Einstein Finch--Skea models demonstrate a remarkable similarity in profile. The density and pressure are monotonically decreasing and the boundary is defined by the pressure vanishing at a finite radius. Since density and pressure are positive everywhere inside the distribution, all the energy conditions are satisfied.

For completeness, the matching with an appropriate exterior spacetime must be achieved.
 In considering the junction conditions  we extrapolate the familiar conditions common in Einstein gravity. However it should be noted that these conditions should be supplemented by more severe boundary conditions in the Lovelock theory as discussed by Davis \cite{davis} regarding the Gauss--Bonnet case.
The vanishing of the pressure at the boundary yields the relationship
\be
\frac{c_1}{c_2} = \frac{(2+CR^2) \tan \sqrt{1+CR^2} -\sqrt{1+CR^2}}{(2+CR^2) + \sqrt{1+CR^2} \tan \sqrt{1+CR^2}} \label{69q}
\ee
between the constants of integration.

Continuity of the gravitational potentials across the boundary $r=R$ also results in the conditions
\beq
e^{-2\lambda (R)} &=& \frac{1}{1+CR^2} = 1-\frac{M}{\sqrt{R}} \label{69u} \\ \n \\
e^{2\nu (R)} &=& 1-\frac{M}{\sqrt{R}}  \n \\
&=& c_1 \left(\frac{3 \sin \sqrt{1+CR^2} - (1+CR^2) \sin \sqrt{1+CR^2} -3\sqrt{1+CR^2}\cos \sqrt{1+CR^2}}{\sqrt{1+CR^2}}\right) \n \\ \n \\
&& + c_2 \left( \frac{(1+CR^2) \cos \sqrt{1+CR^2} -3\cos \sqrt{1+CR^2} -3\sqrt{1+CR^2}\sin \sqrt{1+CR^2}}{\sqrt{1+CR^2}}  \right) \n \\
\label{69w}
\eeq
Eqn (\ref{69u}) immediately gives $C=\frac{M/R^{5/2}}{1-M/\sqrt{R}}$.  Eqns (\ref{69q}) and (\ref{69w}) give
  \be
  c_1 = \frac{\left(1-\frac{M}{\sqrt{R}}\right)VF}{\cos V(GF + HJ)} \hspace{1cm} {\mbox{and}} \hspace{1cm} c_2 = \frac{\left(1-\frac{M}{\sqrt{R}}\right)HV}{\cos V (GF + HJ) }
  \ee
  where we have put $V=\sqrt{1+CR^2}$, $F = (1 +V^2)\tan V - V$, $G=(3-V^2)\tan V - 3V$, $H = (1 +V^2) + V\tan V$ and $J=(V^2 - 3) - 3V\tan V$.
  All the integration constants are now settled explicitly in terms of the mass $M$ and radius $R$ of the perfect fluid distribution. Table 2 illustrates the functional forms for the density and pressure for $d = 4$ and $6$.

\begin{table}
\caption{\bf{Energy density and pressure for $d=2N+2$ ( $N=1, 2$) and $v=\sqrt{1+Cr^2}$ }}

\medskip

\begin{tabular}{|c|c|c|}
  \hline
  & $d =4$ Einstein & $d=6$ GB \\
  \hline
 $\rho$ &$\frac{C(v^2 +2)}{v^4} $ & $ \frac{12 C  (v^2+4)}{v^6} $ \\
 \hline
 $ p$ & $ \frac{C\left( \left(
\zeta-
v \right) \tan v  +\left( \zeta v +1 \right)\right)}{v^2\left(  \left( \zeta+ v \right) \tan v + \left( 1- \zeta
v\right)
\right)}  $ & $  \frac{2C  \left[ (\zeta v - v^2 - 1) \tan v + \zeta (v^2 + 1) + v \right] }{
 v^6 \left[ \left( v^2 - 3 + 3 \zeta v \right) \tan v - \zeta \left( v^2 - 3 \right) + 3 v
 \right]} $ \\
  \hline
\end{tabular}
\end{table}

\end{itemize}

\section{Discussion}

The two canonical fluid distributions, uniform density and isothermal, respectively, approximating to a star interior and a galaxy and a cluster distribution have universality in the gravitational solution. The uniform density sphere is uniquely described \cite{dkm} by the Schwarzschild interior solution in Einstein and Lovelock theory in general in dimensions $ d \geq 4$. So is the case for an isothermal distribution \cite{maha-hans} for pure Lovelock gravity in dimensions $d \geq 2N+2$. A physically acceptable fluid interior lies between these two limiting distributions.  \\

Very recently it has been established \cite{mdk} that a  constant density defines the limiting case of compactness while the other end of the rarest distribution is identified with the Finch-Skea solution. It is interesting that all other bounded distributions are bounded by these two limiting distributions. We have employed these two limiting cases for showing similarity between star solutions in the critical odd and even dimensions $d=2N+1, 2N+2$  for pure Lovelock gravity. In future work, it will be interesting to analyse the behaviour of other well known ansatze in the Lovelock framework. For example, the Tolman solutions will be studied in the Lovelock theory. Now that we have established the pure Lovelock field equations explicitly, it will be possible to study more general Lovelock polynomials in order to examine the corrections of such terms to the standard Einstein theory. It is also a simple matter to supplement the gravitational field equations in pure Lovelock gravity with the Maxwell's equations to discuss the role of charge in stellar distributions.   \\

It has been argued quite strongly \cite{nd, nd2} that gravity has universal behavior in critical odd and even dimensions $d= 2N+1, 2N+2$  in pure Lovelock theory. That means gravitational dynamics in these two critical dimensions should be similar. This then raises the question whether static fluid interiors, like constant density and isothermal distributions, could also have similar behavior in the critical odd and even dimensions. This is the question we had set out to examine in this paper. \\

The master equation to be solved for the fluid distribution is the pressure isotropy equation and it is shown that it has the same character in critical odd and even dimensions $d=2N+1, 2N=2$. It turns out that in $d=2N+1$, there cannot exist any bound distribution meaning that the pressure cannot vanish at any finite radius. This is because pure Lovelock gravity is kinematic in critical odd dimensions $d=2N+1$ \cite{cam-d, bianchi, nd}, which means vacuum is Lovelock flat and hence cannot be an exterior to any distribution. This implies that there cannot exist any bound distribution in critical odd dimensions $d=2N+1$, it has necessarily to be unbounded  cosmological. The pressure isotropy equation reflects this feature. For the critical even dimensions $d=2N+2$ it yields solutions that are similar in their behavior. For this purpose we first obtain a pure GB analogue of the Finch-Skea solution. The similarity between solutions in $d=3, 5$ for odd and $d=4, 6$ for even critical dimensions has been exhibited in Tables I and II. Figs 1-3 also bear testimony to it.   \\

We thus established that for static fluid distributions, pure Lovelock gravitational dynamics is indeed similar in the critical odd and even dimensions $d=2N=1, 2N+2$. \\


{\bf Acknowledgement}
Brian Chilambwe thanks the  DST-NRF Centre of Excellence in Mathematical and Statistical Sciences (CoE-Mass) for support. Opinions expressed and conclusions arrived at are those of the author and are not necessarily to be attributed to the CoE-Mass.

\bibliography{basename of .bib file}

\end{document}